# Providing Higher Throughput for a Single User with M-ary Orthogonal Walsh Codes

Dave Clites, Richard Orr, Jack Rieser, Michael Dellomo
Signal Processing and Communications Analysis Dept.
The MITRE Corporation
McLean, VA 22102
dclites@mitre.org, rorr@mitre.org, jrieser@mitre.org, mdellomo@mitre.org

*Abstract*— Reliable communication is a challenge in a very noisy RF channel further corrupted by severe, multiple narrowband interference. Code Division Multiple Access (CDMA) is a widely used method to both mitigate such interference and support multiple in-band users. The ability to time-control multiple receptions at a receiver permits use of deterministically orthogonal codes. Achieving full capacity of the available frequency band depends, in part, on ensuring that all signals in the band are from the agreed set of orthogonal signals. Each unwanted signal consumes a fraction of the available capacity. In the extreme, as more interferers are added the available bandwidth can support only one user.

We adapt CDMA techniques to a single user by assigning a constellation of *M* symbols built around Walsh codes. This *M*-ary constellation encodes *K* bits of data into one of the length *N* codewords, and gives preference to a non-power-of-two Hadamard length, e.g., 12, 20, and higher multiples. The tradeoff between the modified and conventional Walsh code designs is that one user sends *K* bits per Walsh symbol interval as opposed to many users sending one bit each per symbol. The throughput sacrifice is a consequence of the severe external interference environment.

## I. INTRODUCTION

U.S. Navy sonobuoys use the VHF marine band to transmit their data to an aircraft or ship, in frequency channels 375 kHz wide. The band is also assigned for many other uses, mainly land-based or maritime voice and other narrowband modulations 25 kHz or less. Interference between buoy and other users is limited by line of sight constraints, particularly since sonobuoy operations are typically conducted far enough from land or high density shipping areas to be beyond the horizon of most interferers. As higher altitude receiving aircraft enter service, as sonobuoy deployment areas move toward more littoral regions, and as the number of radios sharing the band increases, the opportunity for interference increases.

Of the available sonobuoy RF channels, each buoy unit operating within line of sight of the receiving aircraft/ship is assigned its own channel, in traditional frequency-division multiple access (FDMA) procedures. As the historical trend continues toward more interference, the channels that can reliably be used for any given buoy will be a shrinking set, to the point that there may not be enough channels available to support a full buoy field. The communications methods used by sonobuoys must adapt to this reality by adding other measures to share the available channels.

There are many potential approaches to dealing with this interference, such as directional receiver antennas or time sharing each of the fewer "good" channels among several units using time-division multiple access (TDMA) network technology. For this effort we focus on another network technology, code-division multiple access (CDMA) methods. This CDMA approach is not intended to singlehandedly resolve rising interference issues, but should be considered along with others for a more complete approach to maintaining reliable sonobuoy communications in future years. Applying network technologies implies the data must be handled in digital form; one current buoy type, AN/SSQ-101 Air Deployable Active Receiver (ADAR), sends digital data instead of the traditional analog-modulated signals.

## II. CDMA WITH MULTIPLE NARROWBAND INTERFERERS

In CDMA, the pairwise cross-correlation functions of the spreading sequences—as seen at a common receiver—characterize mutual interference among received signals. Ideally, all cross-correlations are zero, but this is difficult to realize—especially for mobile terminals—requiring certain coordination among transmitters.

Under time coordination, sequences from different transmitters can be strictly orthogonal at the receiver. For example, Walsh functions—binary sequence sets orthogonal under appropriate alignment—might be used [1]. Absent time coordination, CDMA employs long, statistically orthogonal sequences whose cross-correlations are asymptotically orthogonal with increasing sequence length. During despreading, signals nearly orthogonal to the desired signal become broadband, noise-like waveforms that increase the effective noise power density.

In the latter, Multiple Access Interference (MAI) has virtually negligible impact when the ambient noise density far exceeds total interference power density; otherwise performance deteriorates below that predicted by noise power alone. An unequal distribution of interference power also introduces the near-far problem, see below.

Interference suppression in CDMA is attributable to processing gain. Similarly, CDMA offers substantial advantages in narrowband interference rejection compared to

This work was supported by the Naval Air Systems Command, PMA264, Air Anti-Submarine Warfare Systems Program Office, under contract W15P7T-14-C-A802.

conventional narrowband modulation, i.e., despreading maps narrowband interference into wideband noise. Sufficient processing gain to provide good performance in the presence of both other users and narrowband interference is the goal.

Against interference occupying a significant fraction of the spread bandwidth, however, CDMA can be far less successful, as excessive narrowband interference diminishes several CDMA advantages. This is shown in the general analysis below; also see [1], section 2.7 of [2], and [3]. The analysis assumes a basic knowledge of CDMA principles.

The signal-to-interference-plus-noise ratio (SINR) of a CDMA signal in the presence of thermal noise, other CDMA signals, and multiple narrowband interferers defines potential performance:

$$SINR = E_b/(N_0+I_0). \qquad (1)$$

$E_b$ is energy per bit, $N_0$ is noise power spectral density per Hz, and $I_0$ is composite interference density. $E_b$ is given by $E_b = S/R_b$ where $S$ is signal power at the receiver and $R_b$ is bit rate. In modeling cellular communications, the interference density, $I_0$, is generally broken into three terms,

$$I_0 = I_{sc} + I_{oc} + I_{ni}, \qquad (2)$$

representing interference from: other codes in the same cell ($I_{sc}$), other cells ($I_{oc}$), and narrowband sources ($I_{ni}$). We neglect $I_{oc}$ (assumes only one buoy field transmitting).

$I_{sc}$ depends on parameters of the user signals, but primarily, received signal strength. Not all transmitters necessarily may be at the same distance from the receiver (nor transmit the same power). This near-far problem drove commercial CDMA cellular systems to employ finely tuned power control algorithms that ideally hold received power levels to within ±1 dB. To first order, however, we can assume that local propagation effects, shadowing, and spread of the field are minimal, resulting in roughly constant received signal strength across users.

When the received power $S$ of an individual interfering code is despread over bandwidth $W$, the resultant interference power spectral density is $S/W$. Bandwidth wide enough to capture the entire signal may be written as $W = KR_c$ where constant $K$ (typically ≈1) depends upon both the chip waveform and the statistical distribution of RF phases and chip delays among the received codes. $R_c$ is the CDMA chip rate.

Assuming $N$ active codes within a given cell, $N - 1$ of these are potential interferers with respect to the remaining code. In general, these interferers may not operate constantly, but in the sonobuoy case, each channel may be fully occupied with data, yielding

$$I_{sc} = (N-1)S/W. \qquad (3)$$

$I_{ni}$ is the composite narrowband interference power. Assume individual interferers have average power $S_{ni}$ and bandwidth $W_{ni}$, including rolloff or guard space between adjacent channels (e.g., 25-kHz bandwidth interfering signals with a 10-kHz guard space results in center frequency spacing $W_{ni}$ = 35 kHz). A band thus fully occupied contains $W/W_{ni}$ interferers. Let $p$ be the fraction of the band occupied by narrowband interference; the single-user interference density would be multiplied by $pW/W_{ni}$, yielding total interference

$$I_{ni} = (pW/W_{ni})S_{ni}/W = pS_{ni}/W_{ni}. \qquad (4)$$

For the sonobuoy problem, $p$ can be as high as 0.7 or even 1.0. The impact of this is assessed in the following section.

### III. CDMA CAPACITY IN HIGH INTERFERENCE ENVIRONMENTS

The results of the equations above yield an expression for the effective SINR of the CDMA system. We have

$$\begin{aligned}E_b/(N_0 + I_0) &= E_b/(N_0 + I_{sc} + I_{oc} + I_{ni})\\ &= (S/R)/[N_0 + (N-1)S/W + pS_{ni}/W_{ni}]\\ &= S(W/R)/[WN_0 + (N-1)S + p(W/W_{ni})S_{ni}], \quad (5)\end{aligned}$$

where $R$ is the common user data rate, and the fraction $W/R$ is commonly referred to as the processing gain. We can simplify this equation, first by assuming the number of CDMA users to be sufficiently large that noise thus can be neglected in comparison to CDMA self-interference. Additionally, we temporarily assume absence of external interferers, i.e., $S_{ni} = 0$:

$$SINR = E_b/(N_0+I_0) = (W/R)/(N-1). \qquad (6)$$

Now assume the SINR takes on the minimum threshold value needed for an acceptable error rate (including any margin and losses) so that SINR represents a required value, $SINR_{req}$ as opposed to an achieved value. Solving for $N$ gives:

$$N_{pole} = 1 + (W/R)/SINR_{req} \approx (W/R)/SINR_{re}. \qquad (7)$$

$N_{pole}$ is called the pole capacity of the CDMA system in the absence of external interference; it represents the maximum number of users under the SINR and data rate constraints. The additive 1 shown in (7) is often omitted as a practical compensation for neglected thermal noise. The product $N_{pole}R$ remains constant at fixed SINR and represents the maximum achievable throughput (gross amount of data carried). For example, $SINR_{req} = 0$, $W$ = 400 kHz, $R$ = 25 kbps, imply the system can accommodate 16 users at total net throughput of 400 kbps.

For the realistic sonobuoy problem we cannot, as was done above, ignore external interferers. Rearranging (5) and assuming our equipment's noise figure is low enough to neglect $WN_0$ yields a reduced pole capacity:

$$N = 1 + (W/R)/SINR - [(pW/W_{ni}) S_{ni} + WN_0]/S$$
$$\approx N_{pole} - (pW/W_{ni}) S_{ni}/S. \qquad (8)$$

Equation (8) describes the degradation of pole capacity resulting from the number and total bandwidth of the interferers. For the case of one narrowband interferer, $p = (W_{ni}/W)$ is very small and we find that each interferer degrades pole capacity by $S_{ni}/S$ users. If, for example, $S_{ni}/S$ were between 3 and 6 dB, we would see the pole capacity decline by between 2 - 4 users. Even though the presence of one interferer may cause loss of a few users, this capacity loss is modest when $N_{pole}$ is large. Equation (8) illustrates the resilience of CDMA to a small number of narrowband interferers in showing the fixed-proportion trade-off between the number of interferers tolerated vs. number of users dropped.

Alternately, we can examine the effect of narrowband interferers on the maximum achievable data throughput. Consider a sonobuoy case in which there is only a single user, $N = 1$. Equation (5) can be manipulated to give the maximum tolerable power per external interferer and the upper limit on the user data rate:

$$S_{ni}/S \leq (W_{ni}/R)/[pSIRN_{req}], \text{ and}$$
$$R \leq W_{ni}/[SIRN_{req}/(pS_{ni}/S)]. \qquad (9)$$

Thus, as $S_{ni}$ increases above $S$, and assuming $p$ is small (few narrowband interferers), we not unexpectedly find the achievable bit rate declining inversely with $S_{ni}$. The major impact of narrowband interference upon CDMA systems occurs when interferers begin to occupy a substantial fraction of the band. If $p$ is large, e.g., $p = 0.7$, (8) indicates a steep reduction in pole capacity for terrestrial CDMA, and a marked throughput decline for the sonobuoy case, as shown below:

$$N = 1 + (W/R)/SINR - pW/W_{ni})(S_{ni}/S). \qquad (10)$$

For example, let us assume the $SINR_{req} = 0$ dB, $W_{ni} = 35$ kHz, $W = 400$ kHz and take $p = 0.7$ and $S_{ni} = 2S$. We then have $N = 400/R - 15$. At a data rate $R = 25$ kbps this system supports only a single user, representing a substantial decline in effective throughput from the 400 kbps earlier, when external interference is absent, to 25 kbps.

## IV. CODE SELECTION FOR SINGLE USER SCENARIO

The consequence of the preceding analysis is that in very high interference environments, the benefit of CDMA is reduced to the spreading gain for one user, without meaningful multiple user access. All potential codes are therefore available to the single user, which provide the spreading gain but allow a larger signaling constellation, capable of conveying multiple bits of data with each sequence. There are some practical considerations to guide code length and sequence selection.

Although non-binary symbol sets are possible, the eventual output at the receiver will be a binary data stream. Code selection will result in a constellation of $2^K$ codes, which for this case will be $2^{K-1}$ chip sequences from a Hadamard matrix (of +1 and –1 for binary chip 1 and 0) and their complements. The consequence of this approach is that one user can send $K$ times the traditional CDMA data rate of a single user, instead of many users each sending at the traditional rate. Clearly this is not a good tradeoff in terms of total throughput, so would only be undertaken when the interference is so high that only one user can be supported, based on the preceding analysis. Since $K$ increases logarithmically with code length, this approach is not well suited to $K$ above six to eight bits per symbol. Our testing and analysis examined code lengths 12, 20, 24 and 40, with $K$ values four through six.

In standard form, a Hadamard matrix size $N$ has $N$ columns, $N$ orthogonal rows, with the first row and column all +1's. In the RF environment with high density of narrowband interference, the first code, standard row 1, has a potential problem when used for common FSK or PSK modulations, because a long sequence of the same chip resembles the narrowband interferers and is more difficult to detect in the presence of another NB line near the carrier frequency (for PSK) or the '0' or '1' frequencies at the carrier ± frequency shift (for FSK). Among several ways to avoid this, the preferred method omits the first row and perhaps others with fewer transitions. Leaving out one or more rows means that the typical size $N = 2^M$ (power-of-two) Hadamard matrix will omit half of its possible codes when $2^{M-1}$ are selected. Non-power-of-two Hadamard matrices exist, which for a $2^K$ code selection will achieve the same size constellation as the next larger power of two matrix but with a shorter code length and fewer unused potential codes. An example, matrix $N = 12$ is shown in Table 1, with eight codes (and their complements) selected to form a 16-ary constellation capable of encoding four binary data bits. All the selected codes are balanced between +1 and -1 symbols and for phase modulated signals have no net change in phase over the symbol duration.

TABLE 1: CODE SELECTION FROM $N = 12$ HADAMARD MATRIX

| Bits (complement Bits) | | | | | | | | | | | | |
|---|---|---|---|---|---|---|---|---|---|---|---|---|
| (unused) | +1 | +1 | +1 | +1 | +1 | +1 | +1 | +1 | +1 | +1 | +1 | +1 |
| 0000 (1000) | +1 | -1 | +1 | -1 | +1 | +1 | +1 | -1 | -1 | -1 | +1 | -1 |
| 0001 (1001) | +1 | -1 | -1 | +1 | -1 | +1 | +1 | +1 | -1 | -1 | -1 | +1 |
| 0010 (1010) | +1 | +1 | -1 | -1 | +1 | -1 | +1 | +1 | +1 | -1 | -1 | -1 |
| 0011 (1011) | +1 | -1 | +1 | -1 | -1 | +1 | -1 | +1 | +1 | +1 | -1 | -1 |
| 0100 (1100) | +1 | -1 | -1 | +1 | -1 | -1 | +1 | -1 | +1 | +1 | +1 | -1 |
| 0101 (1101) | +1 | -1 | -1 | -1 | +1 | -1 | -1 | +1 | -1 | +1 | +1 | +1 |
| 0110 (1110) | +1 | +1 | -1 | -1 | -1 | +1 | -1 | -1 | +1 | -1 | +1 | +1 |
| 0111 (1111) | +1 | +1 | +1 | -1 | -1 | -1 | +1 | -1 | -1 | +1 | -1 | +1 |
| (Sync.) | +1 | +1 | +1 | +1 | -1 | -1 | -1 | +1 | -1 | -1 | +1 | -1 |
| (unused) | +1 | -1 | +1 | +1 | +1 | -1 | -1 | -1 | +1 | -1 | -1 | +1 |
| (unused) | +1 | +1 | -1 | +1 | +1 | +1 | -1 | -1 | -1 | +1 | -1 | -1 |

Considering the high level of narrow-band interference, to achieve the required performance likely requires some additional encoding. Our tests used a moderate length Low Density Parity Check (LDPC) code before grouping block-coded bits to select from the *M*-ary symbol codes. Block synchronization is achieved by adding a synchronization pattern to each block. The unused codes from the Hadamard matrix are a natural choice for synchronization because they all (but row one) are net phase neutral and thus support the carrier tracking described above, while being orthogonal to the assigned data codes. Fig. 1 illustrates the phase trajectory of several length-40 Hadamard codes under FSK modulation. It also includes one synchronization code which consists of an unused pattern and its complement at the beginning of each block; this extra length gives 3 dB more gain to the synchronization code and assures the receiver can extract precise timing.

## V. CONCLUSIONS

We note that the calculations used to make the above points rely on several assumptions about the nature of the problem. While details may vary from case to case, the fundamental concerns remain. The principal observations we have found about CDMA may be summarized as follows:

- Statistically orthogonal CDMA systems with high processing gain are reasonably resilient to the mutual interference that arises among the various signals.

- Such CDMA systems are likewise resilient to one or a few individual narrowband interferers whose received power does not substantially exceed that of a CDMA user.

- In the presence of sufficiently many interferers, however, this resilience breaks down and can actually be responsible for decreasing the effective throughput well below that achievable when there is no external interference and either one user or several users deployed on orthogonal codes.

What we have shown is that when the combined noise and interference environment is so severe as to support only a single CDMA user, the set of CDMA codes can be used for data modulation as well as band-spreading, thereby increasing the achievable data rate at a negligible BER trade when error correcting codes such as LDPC are also employed.

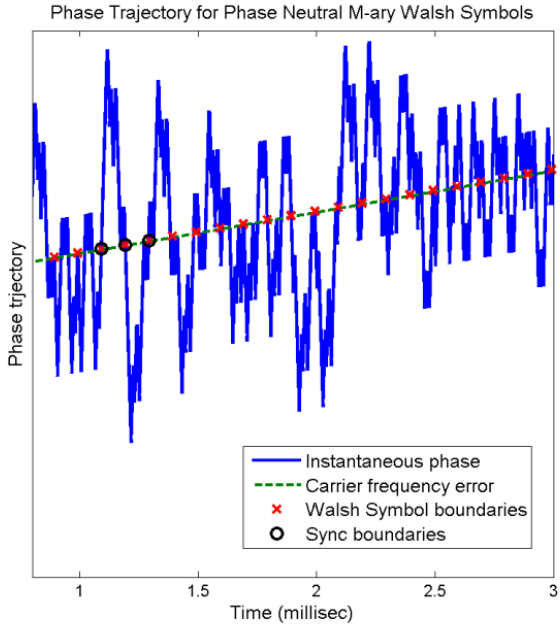

Fig. 1: Phase trajectory and carrier frequency error.